\input{aipcheck}

\documentclass[
    ,final            
  ]
  {aipproc}

\layoutstyle{6x9}

\begin{document}

\title{Suppression of Meszaros' Effect in coupled DE}

\classification{98.80.-k, 98.65.-r}
\keywords      {Cosmology: theory -- dark energy}

\author{Silvio A.~Bonometto}{
  address={Dipart.~di Fisica, Universit\`a di Milano--Bicocca, Piazza della Scienza 3,
 I20126 Milano (Italy) \& INFN, Sezione di Milano--Bicocca},
  email={bonometto@mib.infn.it}
}

\author{Roberto Mainini}{
  email={mainini@mib.infn.it}
}

\begin{abstract}
A phaenomenological DM--DE coupling could indicate their common
origin. Various constraint however exist to such coupling; here we
outline that it can suppress Meszaros' effect, yielding transfered
spectra with a softer bending above $k_{hor,eq}.$ It could be
therefore hard to reconcile these models with both CMB and deep sample
data, using a constant spectral index.
\end{abstract}

\maketitle

A tenable cosmological model must include at least two dark
components, Dark Matter (DM) and Dark Energy (DE); yet only hypotheses
on their nature exist; if DM and DE are physically unrelated, their
presently similar densities are purely accidental.

Attempts to overcome this conceptual deadlock led to suggest, first of
all, that DE has a dynamical nature \cite{DDE}. Interactions between
DM and dynamical DE \cite{interaction},
\cite{interaction2} might then partially cure the problem, keeping close values
for their densities up to large redshift. This option could also be
read as an approach to a deeper reality, whose physical features could
emerge from phenomenological limits to coupling strength and shape.

A longer step forward was attempted by \cite{Mainini2004}, suggesting
that DM and DE are a single complex scalar field, being its quantized
phase and modulus ({\it dual--axion} approach). Instead of introducing
new parameters and limiting them through data fitting, this option,
although cutting the available degrees of freedom, still allows to fit
CMB constraints
\cite{cmb}.

Here, however, we keep on the phenomenological side and discuss
constraints to DM--DE interactions. This will have a fallout also on
the {\it dual--axion} approach, which does face a problem, because of
the feature of the DM--DE coupling it causes.

Any baryon--DE coupling is ruled out by consequences similar to
modifying gravity. Limits are looser for DM--DE coupling, whose
consequences appear only over cosmological distances. Here we outline
a new constraint on this coupling, arising from the early behavior of
fluctuations, over scales destined to evolve into cosmic structures.

Fluctuations over these scales enter the horizon before
matter--radiation equality and their growth is initially inhibited by
the radiative component, then still behaving as a single fluid
together with baryons. While fluctuations in the fluid behave as sonic
waves, DM self gravitation is just a minor dynamical effect is respect
to cosmic expansion. This freezing of fluctuation amplitudes until
equality is known as {\it Meszaros effect}.

\begin{figure}
\includegraphics*[width=8cm,height=8cm]{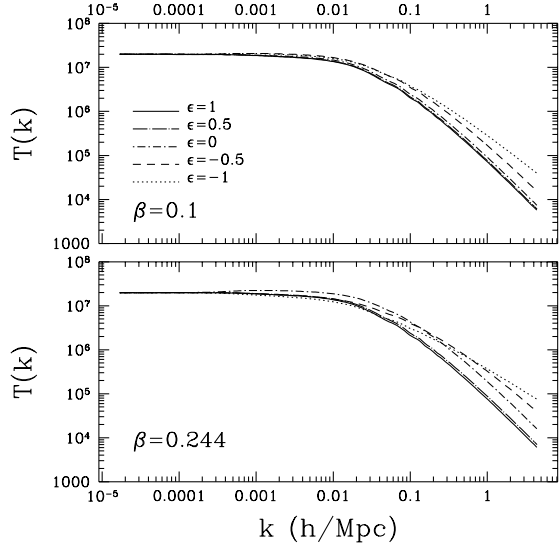}
\caption{Transfer functions for different behaviors of DM--DE coupling
with redshift and/or for different coupling normalization. The case
$\epsilon = 0$ corresponds to redshift independent coupling
intensity. The case $\epsilon = -1$ with $\beta = 0.244$ correspond to
a coupling $C = 1/\phi\, $.  Besides of the different slopes, notice
the dependence on the model of the bending scale and, in particular,
its dependence on the coupling strength, also in constant coupling
models (dash--dotted lines).  }
\label{transfer}
\end{figure}
\begin{figure}
\includegraphics*[width=8cm,height=6.cm]{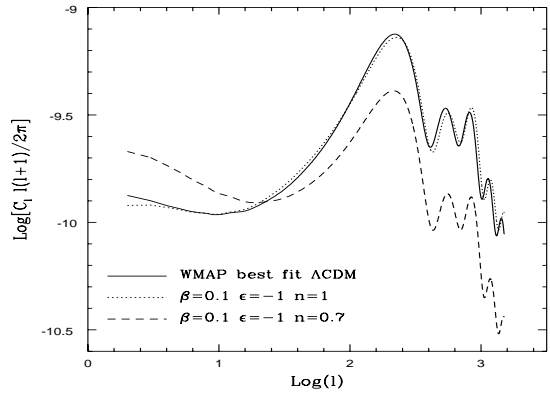}
\caption{Anisotropy spectrum of the $\Lambda$CDM model yielding the
best fit to WMAP3 data compared with the spectra for coupled models
with $\beta = 0.1$ and $\epsilon = 1,$ for $n = 1 $ and $n = 0.7.$ In
the former case one can expect to recover a reasonable fit to data by
adjusting other model parameters. Taking $n = 0.7,$ a value just
acceptable to fit deep sample data, any fitting to CMB anisotropy data
is apparently excluded.  }
\label{Cl}
\end{figure}
\begin{figure}
\includegraphics*[width=9cm,height=6cm]{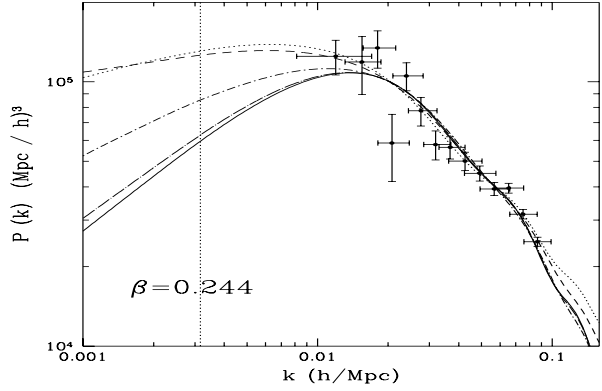}
\caption{Comparison with SDSS digital survey data.  Different
curves refer to different $\epsilon $'s with the solid line ($\epsilon
= 1$) essentially coinciding with an uncoupled model. Constant
coupling models ($\epsilon = 0$) are described by dot--dashed
curves. Negative $\epsilon$'s yield a further decrease of $n$. The
vertical dotted line is the approximate scale where the Sachs \& Wolfe
$C_l$ plateau begins. }
\label{fitbeta2}
\end{figure}
\begin{figure}
\includegraphics*[width=9cm,height=7cm]{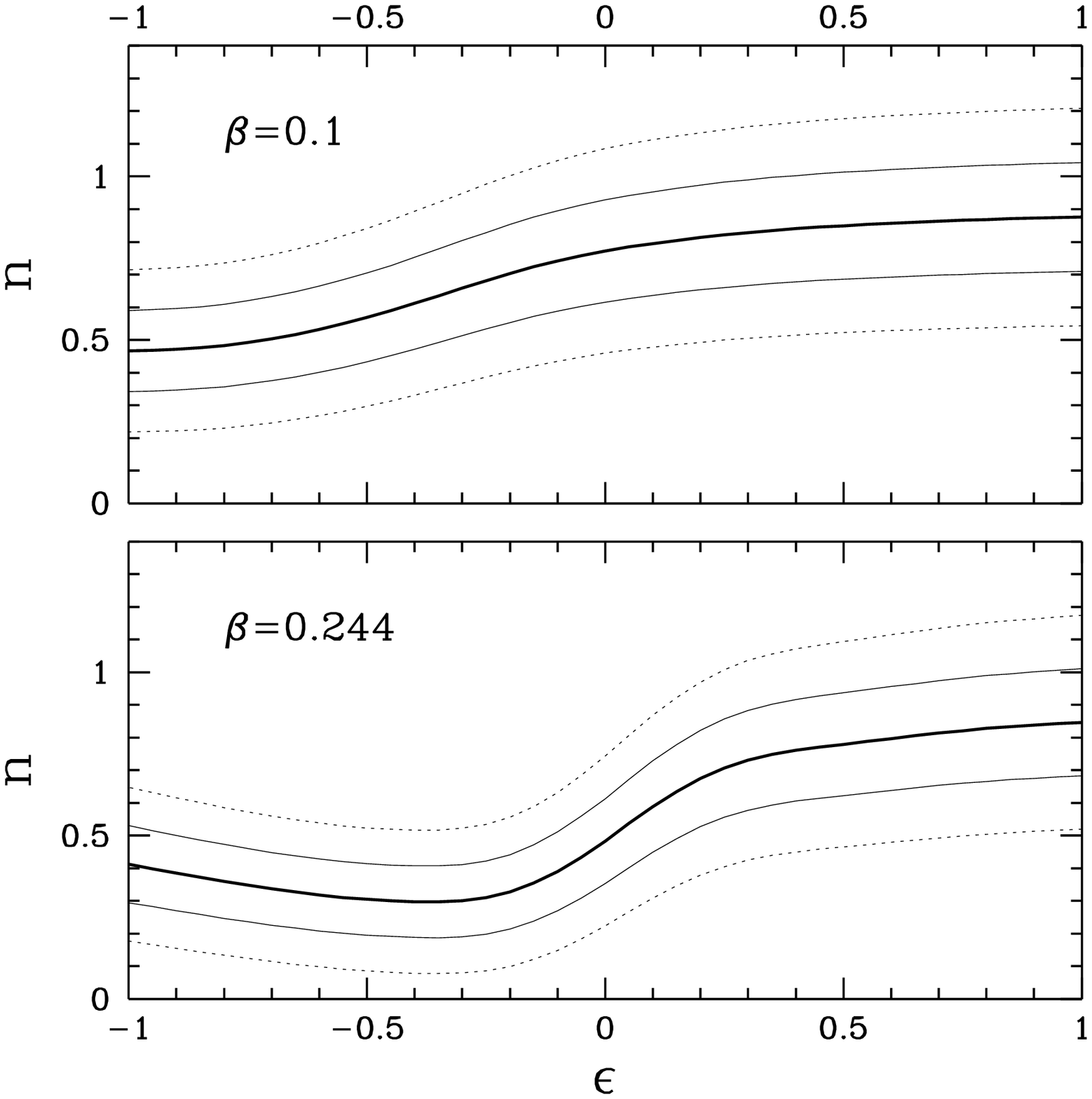}
\caption{1-- and 2--$\sigma$ intervals of $n$, when $\epsilon$ varies,
for the $\beta$ values in the frames}
\label{ennealfa}
\end{figure}
Here we wish to outline that DM--DE coupling can damp Meszaros effect,
so that fluctuation growth, between the entry in the horizon and
equality, is significantly enhanced. As a matter of fact, fluctuation
freezing is essential, in shaping the transfered spectrum.

The freezing or its damping only marginally affect baryons and
radiation. Therefore, while the transfer function changes, CMB spectra
keep almost unaffected. All details on the way how these results are
obtainable can be found in \cite{P1}. Let us however outline that what
we wish to outline is a rather major effect, which allows to discard a
class of models, {\it a priori}; no general data fitting, constraining
parameters and/or showing specific model advantages, needs then to be
performed here.

Accordingly, we keep to cosmological parameter values ensuing from
WMAP3 best--fit \cite{Spergel2006}, although deduced by assuming a
$\Lambda$CDM model. In particular, we shall take $\Omega = 1$; the
present value of the cold DM (baryon) density parameter will be
$\Omega_{o,c} = 0.224$ ($\Omega_{o,b} = 0.044$); the Hubble parameter
will be $h = 0.704.$

Our analysis here will however be restricted to SUGRA 
potentials \cite{Brax}
\begin{equation}
\label{ptntl}
V(\phi) = (\Lambda^{\alpha+4}/\phi^\alpha) \exp(4 \pi \phi^2/m_p^2)
\end{equation}
with $\Lambda = 100~$GeV. Here $m_p = G^{-1/2}$ is the Planck mass.

The background equations for coupled DE and DM, using the conformal
time $\tau,$ read
\begin{equation}
\ddot{\phi} + 2(\dot a/ a) \dot{\phi} + a^2 V_{,\phi} = 
C(\phi) a^{2} \rho _{c}~, ~~~~ \dot{\rho _{c}} + 3 (\dot a /
a)\rho_{c} = - C(\phi) \dot \phi \rho _{c}~
\label{backg}
\end{equation}
where the coupling is set by
\begin{equation}
C(\phi) = 4 \sqrt{\pi/ 3} (\beta / m_p)
\left( \phi / m_p \right)^\epsilon
= 4 \sqrt{\pi/ 3} (\tilde \beta / m_p)
\label{C}
\end{equation}
and therefore parametrized by $\beta,$ $\epsilon$ or $\tilde \beta.$
The {\it dual--axion} model naturally predicts a coupling $C=1/\phi$,
consistent with eq.~(\ref{C}) if $\epsilon = -1$ and $\beta \simeq
0.244. $

Transfer functions are shown in Fig.~\ref{transfer} for $\beta= 0.1$
and 0.244 and various~$\epsilon$'s.

The suppression of fluctuation freezing is obviously stronger for
greater $\beta $ (and increasingly negative $\epsilon$ values). For
$\epsilon = -1$, enclosing the case $C = 1/\phi$ when $\beta = 0.244$,
the steepness of the transfer function, for $k > k_{hor,eq}$ is much
reduced.  The effect is still significant also for $\epsilon = -0.5,$
namely when~$\beta = 0.244\, .$ 

While this occur, the CMB anisotropy spectum keeps a regular behavior,
as is shown in Figure~\ref{Cl}; here we compare WMAP3 data with $C_l$
for $\beta = 0.1$, $\epsilon = -1 $ and $n = 1$ or 0.7~.
Transfered spectra, compared with SDSS data (see Fig.~\ref{fitbeta2}), 
show in fact that $n \sim 0.5$--0.7 is needed to fit data.

In the case $\epsilon \neq 0,$ the discrepancy from unity of the
spectral index $n,$ assumed to be constant, is a measure of the
distortion caused by the suppression of Meszaros effect.

If we take $n=0.85$ at 1--$\sigma $ as a threshold to discard a model,
no $\epsilon < 0$ model~is~al\-lowed with $\beta = 0.244,$ while
$\epsilon < -0.16$ are also inhibited with $\beta = 0.1\, $. At
2--$\sigma$'s~the~situ\-ation is not much improved for $\beta = 0.244$,
while lower values of $\epsilon$ are admitted for $\beta = 0.1\, $.

In particular, $C=1/\phi$, as for the {\it dual--axion} model, is
outside the range indicated.  Modification to make this model
consistent with data were however proposed \cite{last}.

The analysis was extended to models with positive $\epsilon,$ for
which coupling rises while $\phi$ increases. A large deal of these
models is apparently allowed.


{}

\end{document}